\documentclass[aps,prb,reprint,twocolumn,groupedaddress,showpacs]{revtex4}

\usepackage{amsmath}
\usepackage{amsfonts}
\usepackage{mathrsfs}

\usepackage{epsfig}

  \usepackage{natbib}

\begin{document}

\newcommand{\bra}[1]{\langle #1|}
\newcommand{\ket}[1]{|#1\rangle}
\newcommand{\braket}[2]{\langle #1|#2\rangle}

\newcommand{\onehalf}[0]{\frac{1}{2}}  
\newcommand{\overroot}[0]{\frac{1}{\sqrt 2}}  
\newcommand{\halfDelta}[0]{\frac{\Delta}{2}}  
\newcommand{\half}[1]{\frac{#1}{2}}  


\title{Quantum Tunneling of the Magnetic Moment in a Free Particle}


\author{M. F. O'Keeffe,  E. M. Chudnovsky, and D. A. Garanin}
\affiliation{Physics Department, Lehman College, City University
of New York,
    250 Bedford Park Boulevard West, Bronx, New York, 10468-1589, USA}

\date{\today}

\begin{abstract}
We study tunneling of the magnetic moment in a particle that has
full rotational freedom. Exact energy levels are obtained and the
ground-state magnetic moment is computed for a symmetric rotor.
The effect of the mechanical freedom on spin tunneling manifests
itself in a strong dependence of the magnetic moment on the
moments of inertia of the rotor. Energy of the particle exhibits
quantum phase transitions between states with different values of
the magnetic moment. Particles of various shapes are investigated
and quantum phase diagram is obtained.
\end{abstract}


\maketitle



\section{Introduction \label{sec:intro}}
%
There has been much recent interest in quantum mechanics of
nanoscopic magnets that possess mechanical freedom. Experimental
work in this area focused on free magnetic clusters
\cite{Billas,Xu,Payne}, magnetic particles that are free to move
inside solid nanocavities \cite{Tejada}, magnetic microresonators
\cite{NIST,Davis}, and magnetic molecules bridged between
conducting leads \cite{Heersche,Henderson,Voss}. Theoretical
research on free magnetic particles has been scarce. The generic
problem is that of a rigid quantum rotor with a spin. Without a
spin this problem is tractable by analytical methods only for a
symmetric rotor \cite{Edmonds}. Complications resulting from spin
degrees of freedom make even symmetric cases significantly more
difficult \cite{VanVleck}.

First attempt to understand how mechanical freedom of a small
magnetic particle affects tunneling of the magnetic moment was
made in Ref. \onlinecite{EC-PRL94}, where it was noticed that
tunneling of a macroscpin in a free particle must be entangled
with mechanical rotations in order to conserve the total angular
momentum (spin + orbital). Similar situation occurs for tunneling
of a superconducting current between clockwise and
counterclockwise directions in a SQUID
\cite{Friedman-SQUID,Wal-SQUID,Clarke-2008}. Recently, it was
demonstrated \cite{chugar10prb} that the problem of a rigid rotor
with a spin can be solved exactly in the laboratory frame when
mechanical rotation is allowed only about a fixed axis and the
spin states are reduced to spin-up and spin-down due to strong
magnetic anisotropy. The latter is typical for magnetic molecules
and small ferromagnetic clusters \cite{EC-JETP,EC-Gunther,MQT}.
The reduction to two spin states in a system rotating about a
fixed axis also allows one to obtain simple solution of the
problems of a magnetic molecule embedded in a microcantilever
\cite{Reem-PRL}, magnetic molecule vibrating between conducting
leads \cite{Reem-EPL}, and of a macrospin tunneling inside a
torsional resonator \cite{Kovalev,Gar-EC-2011}. However, the
problem for arbitrary rotations of a two-state spin system, which
is relevant to free magnetic nanoparticles, remained unsolved
until now.

In this paper we show that the problem of a two-state macrospin
inside a symmetric rigid rotor has rigorous solution for arbitrary
rotations in the coordinate frame that is rigidly coupled to the
rotor. Magnetic moment of the electrically neutral rotor is
entirely due to spin. It depends on the relative contribution of
the up and down spin states. When a nanoparticle is embedded in a
solid, tunneling of the spin results in a zero ground-state
magnetic moment. This situation changes for a free particle due to
a complex interplay between spin and mechanical angular momentum
that conserves the total angular momentum. We show that the energy
of the particle exhibits first- or second-order quantum phase
transitions between states with different values of the total
angular momentum when the latter is treated as a continuous
variable. The order of the transition depends on the shape of the
particle. The ground-state magnetic moment of a free particle with
a total spin $S$ can be anything between zero and $g{\mu_B}S$,
depending on the principal moments of inertia (with $\mu_B$ being
the Bohr magneton and $g$ being the gyromagnetic factor associated
with the spin).

The structure of the article is as follows. Quantum theory of a
rigid rotator is briefly reviewed in Sec. \ref{rotation}. Theory
of a tunneling macrospin is reviewed in Sec. \ref{tunneling}.
Quantum states of a rigid rotator containing a tunneling macrospin
are constructed in Sec. \ref{rotor-spin}. Ground state of a
symmetric rotor with a spin is analyzed in Sec. \ref{ground}.
Ground-state magnetic moment is studied in Sec. \ref{moment}. Our
conclusions are presented in Section \ref{conclusions}.


\section{Quantization of rigid body rotations \label{rotation}}

Consider first the problem without a spin. We choose the
coordinate frame that is rigidly coupled with the rotating body
and direct the axes of that frame $x, y$, and $z$ along the
principle axes of the tensor of moments of inertia of the body. In
such coordinate frame the Hamiltonian of mechanical rotations is
given by~\cite{Edmonds}
%
\begin{equation}
\hat H_R = \frac{\hbar^2}{2} \left ( \frac{L_x^2}{I_x} +
\frac{L_y^2}{I_y} + \frac{L_z^2}{I_z} \right )\,.
\label{eq:rotation_hamiltonian}
\end{equation}
Here $I_x, I_y, I_z$ are the principal moments of inertia and
$L_x, L_y, L_z$ are projections of the operator of the mechanical
angular momentum, defined in the fixed laboratory coordinate
frame, onto the body axes $x, y, z$.  Such a choice of coordinates
and operators results in the anomalous  commutation relations
\cite{Klein}, $[L_i, L_j] = - i \epsilon_{ijk} L_k$ (notice the
minus sign in the right-hand side), but does not affect the
relations ${\bf L}^2 = L(L+1)$, $[{\bf L}^2,L_z]=0$.

For a symmetric rotor two of the moments of inertia are the same,
$I_x = I_y$, and the Hamiltonian can be written as
\begin{equation}
\hat H_R =
    \frac{\hbar^2 \mathbf L^2}{2 I_x} + \frac{\hbar^2 L_z^2}{2}
    \left ( \frac{1}{I_z} - \frac{1}{I_x} \right )\,.
\label{H-L}
\end{equation}
The corresponding eigenstates are characterized by three quantum
numbers $L, K$, and $M$,
\begin{align}
    \mathbf L^2 \ket{LKM} &= L(L+1) \ket{LKM},  \; L = 0, 1, 2, \ldots  \nonumber  \\
        L_z \ket{LKM} &= K \ket{LKM}, \; K = -L, -L+1,  \ldots, L-1, L  \nonumber  \\
        L_Z \ket{JKM} &= M \ket{LKM}, \; M = -L, -L+1\ldots, L-1,
        L\,,
\end{align}
where $L_Z$ is the angular momentum operator defined with respect
to the laboratory coordinate frame ($X,Y,Z$). The eigenvalues of
(\ref{H-L}) are degenerate on $M$:
\begin{equation}
E_{LK} = \frac{\hbar^2 L(L + 1)}{2 I_x} + \frac{\hbar^2 K^2}{2}
\left ( \frac{1}{I_z} - \frac{1}{I_x} \right )\,.
\label{eq:symmetric_rotor_energy}
\end{equation}

The general form for the energy levels of a rotating asymmetric
rigid body, $I_x \neq I_y \neq I_z$, does not exist, although it
is possible to calculate matrix elements of the Hamiltonian for a
given $L$.


\section{Tunneling of a large spin \label{tunneling}}

Let ${\bf S}$ be a fixed-length spin  embedded in a stationary
body. Naturally, the magnetic anisotropy is defined with respect
to the body axes. The general form of the crystal field
Hamiltonian is
\begin{equation}
\hat H_S = \hat H_\parallel + \hat H_\perp\,,
\end{equation}
where $\hat H_\parallel$ commutes with $S_z$ and $\hat H_\perp$ is
a perturbation that does not commute with $S_z$.  The states
$\ket{\pm S}$ are degenerate ground states of $\hat H_\parallel$,
where $S$ is the total spin of the nanomagnet.  $\hat H_\perp$
slightly perturbs these states, adding to them small contributions
from other $\ket{m_S}$ states.  We will call these degenerate
perturbed states $\ket{\psi_{\pm S}}$.  Physically they describe
the magnetic moment aligned in one of the two directions along the
anisotropy axis.  Full perturbation theory with account of the
degeneracy of $\hat H_S$ provides quantum tunneling between the
$\ket{\psi_{\pm S}}$ states for integer $S$.  The ground state and
first excited state are symmetric and antisymmetric combinations
of $\ket{\psi_{\pm S}}$, respectively \cite{MQT},
\begin{align}
\Psi_+ &= \overroot (\ket{\psi_S} + \ket{\psi_{-S}}) \nonumber  \\
\Psi_- &= \overroot (\ket{\psi_S} - \ket{\psi_{-S}})\,,
\label{eq:superpos}
\end{align}
which satisfy
\begin{equation}
\hat H_S \Psi_\pm = E_\mp \Psi_\pm\,,
\end{equation}
where
\begin{equation}
E_+ - E_- \equiv \Delta.
\end{equation}

The tunnel splitting $\Delta$ is generally many orders of
magnitude smaller than the distance to other spin energy levels,
which makes the two-state approximation very accurate at low
energies.  For example,
\begin{equation}
\hat H_S = - D S_z^2 + d S_y^2 \label{CF}
\end{equation}
with $d \ll D$ describes the biaxial anisotropy of spin-10
molecular nanomagnet Fe-8, where the tunnel splitting in the limit
of large $S$ is given by\cite{Garanin:1991st}
\begin{equation}
\Delta = \frac{8 S^{3/2}}{\pi^{1/2}} \left ( \frac{d}{4 D} \right
)^S D. \label{Delta}
\end{equation}
The distance to the next excited spin level is $(2 S - 1) D$,
which is large compared to $\Delta$.

It is convenient to describe these lowest energy spin states
$\Psi_\pm$ with a pseudospin-$1/2$. Components of the
corresponding Pauli operator $\boldsymbol \sigma$ are
\begin{align}
\sigma_x &= \ket{\psi_{-S}} \bra{\psi_{S}} + \ket{\psi_{S}} \bra{\psi_{-S}}  \nonumber  \\
\sigma_y &= i \ket{\psi_{-S}} \bra{\psi_{S}} - i \ket{\psi_{S}} \bra{\psi_{-S}}  \nonumber  \\
\sigma_z &= \ket{\psi_{S}} \bra{\psi_{S}} - \ket{\psi_{-S}}
\bra{\psi_{-S}}\,.
\end{align}
The projection of $\hat H_S$ onto $\ket{\psi_{\pm S}}$ states is
%
\begin{equation}
\hat H_\sigma = \sum_{m, n = \psi_{\pm S}} \bra{m} \hat H_S
\ket{n} \; \ket{m} \bra{n}. \label{eq:project}
\end{equation}
Expressing $\ket{\psi_{\pm S}}$ in terms of $\Psi_\pm$ one obtains
\begin{equation}
\bra{\psi_{\pm S}} \hat H_S \ket{\psi_{\pm S}} = 0, \qquad
    \bra{\psi_{\pm S}} \hat H_S \ket{\psi_{\mp S}} = - \halfDelta\,,
\end{equation}
which gives the two-state Hamiltonian
\begin{equation}
\hat H_\sigma = - \halfDelta \sigma_x
\end{equation}
having eigenvalues $\pm \Delta/2$.

In the absence of tunneling a classical magnetic moment is
localized in the up or down state. It is clear that delocalization
of the magnetic moment due to spin tunneling reduces the energy by
$\Delta/2$. In a free particle, however, tunneling of the spin
must be accompanied by mechanical rotations in order to conserve
the total angular momentum. Such rotations cost energy, so it is
not a priori clear whether the tunneling will survive in a free
particle and what the ground state is going to be. This problem is
addressed in the following Section.
%


\section{Rigid rotor containing tunneling macrospin \label{rotor-spin}}

Consider now a tunneling macrospin embedded in a free particle
having the body $z$-axis as the magnetic anisotropy direction.
Such a particle is characterized by the total angular momentum,
${\bf J} = {\bf L} + {\bf S}$. In the body frame this operator may
appear unconventional due to the different sign of commutation
relations for ${\bf L}$ and ${\bf S}$. However, this problem can
be easily fixed \cite{VanVleck} by the transformation ${\bf S}
\rightarrow - {\bf S}$ that changes the sign of the commutation
relation for ${\bf S}$. Such a transformation does not change the
results of the previous Section because the crystal field
Hamiltonian contains only even powers of ${\bf S}$. It is
interesting to notice that while in the laboratory frame
$[J_i,S_j] = i\epsilon_{ijk}S_k$, components of the operators
${\bf J}$ and ${\bf S}$ defined in the body frame commute with
each other \cite{VanVleck}. In addition, operator ${\bf J}^2$ is
the same in the body and laboratory frames \cite{Klein}. This
permits description of quantum states of the particle in terms of
quantum numbers associated independently with the total angular
momentum and spin.

The full Hamiltonian is given by the sum of the rotational energy
and magnetic anisotropy energy
\begin{equation}
\hat H = \frac{\hbar^2 L_x^2}{2 I_x} + \frac{\hbar^2 L_y^2}{2 I_y}
+ \frac{\hbar^2 L_z^2}{2 I_z}
        + \hat H_S\,.
\end{equation}
Expressing the mechanical angular momentum $\mathbf L$ in terms of
the total angular momentum $\mathbf J$ and the spin $\mathbf S$,
we get
\begin{align}
\hat H  &= \frac{\hbar^2}{2} \left ( \frac{J_x^2}{I_x} +
\frac{J_y^2}{I_y} + \frac{J_z^2}{I_z} \right ) +
        \frac{\hbar^2}{2} \left ( \frac{S_x^2}{I_x} + \frac{S_y^2}{I_y} + \frac{S_z^2}{I_z} \right )
            \nonumber  \\
        & \qquad -\hbar^2 \left ( \frac{J_x S_x}{I_x} + \frac{J_y S_y}{I_y} + \frac{J_z S_z}{I_z} \right )
        + \hat H_S\,.
\label{eq:full_hamiltonian}
\end{align}
For a symmetric rigid rotor with $I_x = I_y$ this Hamiltonian
reduces to
%
\begin{eqnarray}\label{eq:hamiltonian-sym}
\hat H & = &
    \frac{\hbar^2 \mathbf J^2}{2 I_x} + \frac{\hbar^2 J_z^2}{2} \left ( \frac{1}{I_z} -
    \frac{1}{I_x} \right ) \nonumber \\
    & -  & \hbar^2 \left ( \frac{J_x S_x+J_y S_y}{I_x}\,  + \frac{J_z S_z}{I_z} \right )
        + \hat H_S'\,,
\end{eqnarray}
where
\begin{equation}\label{H'}
\hat H_S' = \hat H_S + \frac{\hbar^2}{2}\left(\frac{1}{I_z} -
\frac{1}{I_x}\right)S_z^2 + \frac{\hbar^2 {\bf S}^2}{2I_x}\,.
\end{equation}
The last term in $\hat H_S'$ is an unessential constant,
$\hbar^2S(S+1)/(2I_x)^2$. The second term provides renormalization
of the crystal field in a freely rotating particle. For, e.g., the
biaxial spin Hamiltonian given by Eq.\ (\ref{CF}) it leads to
\begin{equation}\label{D'}
D \rightarrow D - \frac{\hbar^2}{2}\left(\frac{1}{I_z} -
\frac{1}{I_x}\right)\,.
\end{equation}
This, in turn, renormalizes the tunnel splitting given by Eq.\
(\ref{Delta}). For a particle that is allowed to rotate about the
$Z$-axis only (that is, in the limit of $I_x \rightarrow \infty$)
these results coincide with the results obtained by the instanton
method in Ref.\ \onlinecite{okechu11prb}, where it was shown that,
in practice, the renormalization of the magnetic anisotropy and
spin tunnel splitting by mechanical rotations is small. Eq.\
(\ref{D'}) provides generalization of this effect for arbitrary
rotations of a symmetric rotator with a spin. According to this
equation and Eq.\ (\ref{CF}), when rotations are allowed the
effective easy-axis magnetic anisotropy and the tunnel splitting
can decrease or increase, depending on the ratio $I_x/I_z$.

Projection of Eq.\ (\ref{eq:hamiltonian-sym}) on the two spin
states along the lines of the previous Section gives
%
\begin{equation}
\hat H =
    \frac{\hbar^2 \mathbf J^2}{2 I_x} + \frac{\hbar^2 J_z^2}{2} \left ( \frac{1}{I_z}
    - \frac{1}{I_x} \right )
     - \halfDelta \sigma_x
    - \frac{\hbar^2 S}{I_z} J_z \sigma_z\,.
\label{eq:hamiltonian}
\end{equation}
where we have used
\begin{equation}
    \bra{\psi_{\pm S}} S_z \ket{\psi_{\pm S}} = \pm S\,,
    \qquad \bra{\psi_{\pm S}} S_{x,y} \ket{\psi_{\pm S}} = 0\,.
\end{equation}
We construct eigenstates of this Hamiltonin according to
%
\begin{equation}
    \ket{\Psi_{JK}} = \overroot (C_{\pm S} \ket{\psi_S} \pm C_{\mp S} \ket{\psi_{-S}} )
    \ket{JK}
    \label{eq:eigenstates}
\end{equation}
where
\begin{align}
    \mathbf J^2 \ket{JK} &= J(J+1) \ket{JK},  \; J = 0, 1, 2, \ldots  \nonumber  \\
        J_z \ket{JK} &= K \ket{JK}, \; K = -J, \ldots, J.
\end{align}
Solution of $\hat H  \ket{\Psi_{JK}} = E \ket{\Psi_{JK}}$ gives
energy levels as
\begin{equation}
E_{JK}^{(\pm)}=E_{JK}
\pm\sqrt{\left(\frac{\Delta}{2}\right)^{2}+\left(\frac{\hbar^{2}KS}
{I_{z}}\right)^{2}}\,, \label{eigenstates-EJK}
\end{equation}
where $E_{JK}$ is provided by Eq.\
(\ref{eq:symmetric_rotor_energy}) with $L$ replaced by $J$. The
upper (lower) sign in Eq.\ (\ref{eigenstates-EJK}) corresponds to
the lower (upper) sign in Eq.~\eqref{eq:eigenstates}. For $K \neq
0$ each state is degenerate with respect to the sign of $K$.  For
$K = 0, 1, 2, \ldots$ the coefficients in Eq.\
(\ref{eq:eigenstates}) are given by
%
\begin{equation}
C_{\pm} = \sqrt{ 1 \pm \alpha K / \sqrt{ S^2 + (\alpha K)^2 }}\,,
\label{eq:C}
\end{equation}
where $\alpha$ is a dimensionless magneto-mechanical ratio,
\begin{equation}
\alpha = \frac{2 (\hbar S)^2}{I_z \Delta} \label{eq:alpha}\,.
\end{equation}

Energy levels in Eq.\ (\ref{eigenstates-EJK}) can be given a
simple semiclassical interpretation. Indeed, the last term in this
equation is the tunnel splitting of the levels in the effective
magnetic field that appears in the body reference frame due to
rotation about the spin quantization axis at the angular velocity
$\hbar K/I_z$. When $S = 0$ (which also means $\Delta = 0$)
Eq.~\eqref{eigenstates-EJK} with $J = L$ gives the energy of the
quantum symmetric rigid rotor without a spin,
Eq.~\eqref{eq:symmetric_rotor_energy}. In the case of a heavy body
(large moments of inertia) the ground state and the first excited
state correspond to $J = K = 0$, and we recover the tunnel-split
spin states in a non-rotating macroscopic body, $E_{00\pm} = \pm
\Delta/2$. In the general case, spin states of the rotator are
entangled with mechanical rotations.

Equations (\ref{eq:eigenstates})-(\ref{eq:C}) are our main
analytical results for the low-energy states of a free magnetic
particle. In general, numerical analysis is needed to find the
ground state of the particle. Special cases of the aspect ratio
that will be analyzed below include a needle of vanishing diameter
(which is equivalent to the problem of the rotation about a fixed
axis treated previously in the laboratory frame by two of the
authors \cite{chugar10prb}), a finite-diameter needle, a sphere,
and a disk.


\section{Ground state  \label{ground}}
Minimization of the energy in Eq.\ (\ref{eigenstates-EJK}) on $J$
with the account of the fact that $J$ cannot be smaller than $K$
immediately yields $J = K$, that is, the ground state always
corresponds to the maximal projection of the total angular
momentum onto the spin quantization axis. In semiclassical terms
this means that the minimal energy states in the presence of spin
tunneling always correspond to mechanical rotations about the
magnetic anisotropy axis. This is easy to understand by noticing
that the sole reason for mechanical rotation is the necessity to
conserve the total angular momentum while allowing spin tunneling
to lower the energy. To accomplish this the particle needs to
oscillate between clockwise and counterclockwise rotations about
the spin quantization axis in unison with the tunneling spin. If
such mechanical oscillation costs more energy than the energy gain
from spin tunneling, then both spin tunneling and mechanical
motion must be frozen in the ground state as, indeed, happens in
very light particles (see below). Rotations about axes other than
the spin quantization axis can only increase the energy and, thus,
should be absent in the ground state.
\begin{figure}
    \includegraphics[scale=0.32, angle=-90]{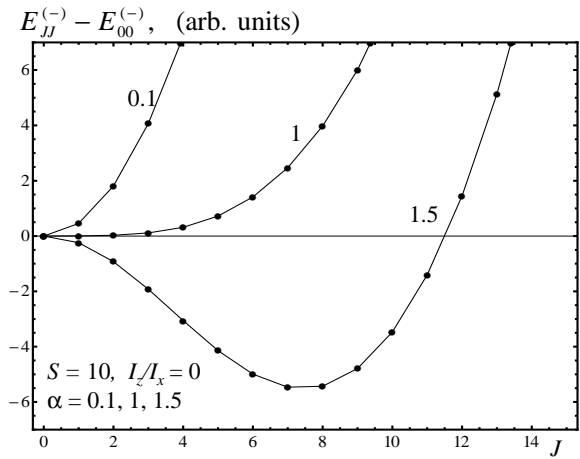}
\caption{Dependence of energy on $J$ at $K=J$ and $I_z/I_x = 0$
for different values of $\alpha$. The plot shows second-order
quantum phase transition on $\alpha$.} \label{Fig-EJJ_lambda=0}
\end{figure}
\begin{figure}
    \includegraphics[scale=0.32, angle=-90]{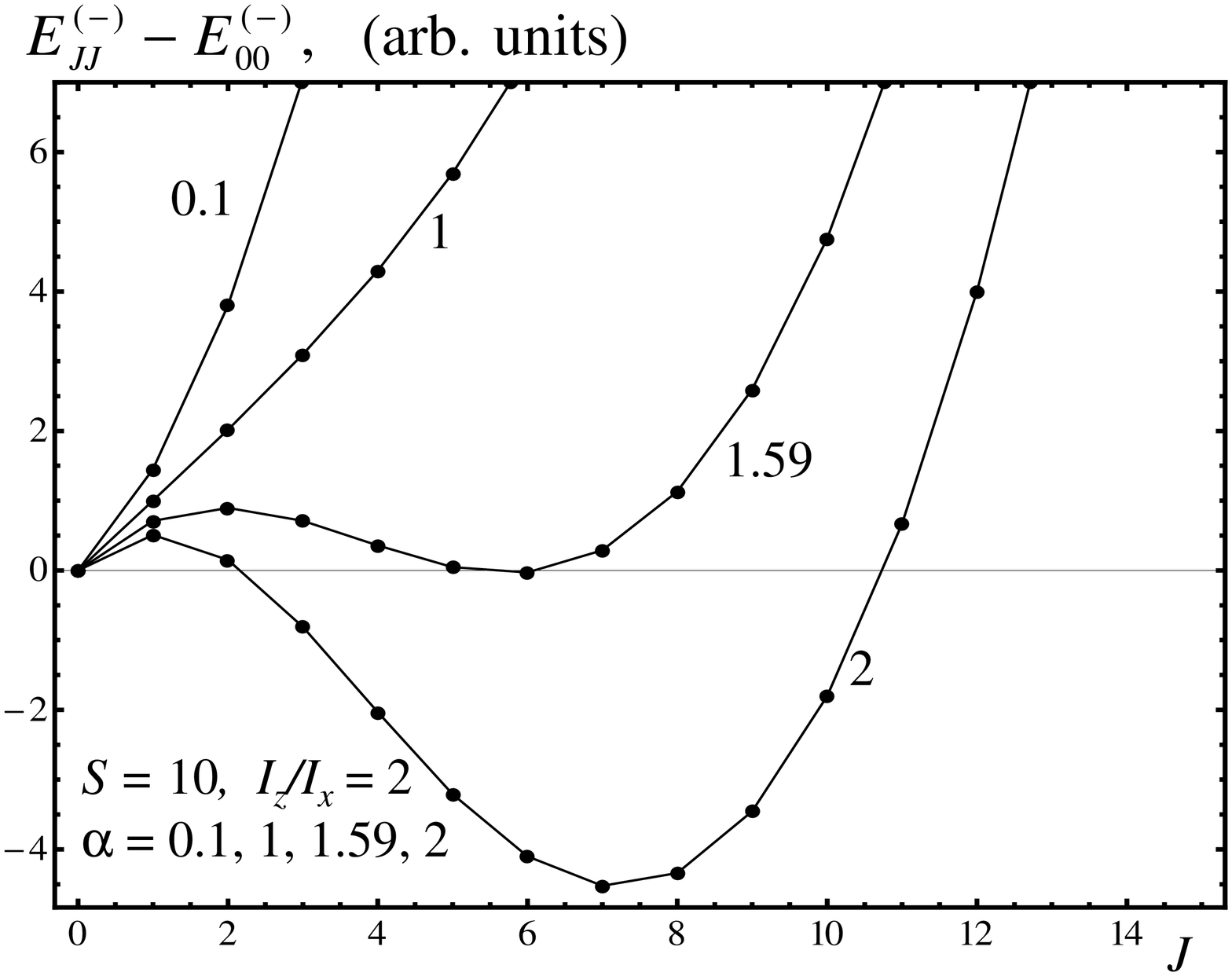}
\caption{Dependence of energy on $J$ at $K=J$ and $I_z/I_x = 2$
for different values of $\alpha$. The plot shows first-order
quantum phase transition on $\alpha$.} \label{Fig-EJJ_lambda=2}
\end{figure}

For further analysis it is convenient to write Eq.\
(\ref{eigenstates-EJK}) in the dimensionless form,
\begin{equation}
\frac{E_{JK}^{(\pm)}}{\Delta} = \frac{\alpha}{4} \Bigg[\frac{ J
(J+1) - K^2 }{S^2} \lambda + \frac{K^2}{S^2} \Bigg ]
    \pm \frac{1}{2}\sqrt{1 + \frac{K^2}{S^2} \alpha^2} \,,
\label{eq:energy}
\end{equation}
in terms of dimensionless parameters $\alpha$  and the aspect
ratio for the moments of inertia
\begin{equation}
\lambda = {I_z}/{I_x}\,. \label{eq:lambda}
\end{equation}
The range of $\lambda$ for a symmetric rotator is $0 \leq \lambda
\leq 2$. For, e.g., a symmetric ellipsoid with semiaxes $a=b \neq
c$, one has $\lambda = 2a^2/(a^2 + c^2)$.

The dependence of the energy levels (\ref{eigenstates-EJK}) on $J$
at $K = J$ is shown in figures \ref{Fig-EJJ_lambda=0} and
\ref{Fig-EJJ_lambda=2}. It exhibits quantum phase transition on
the parameter $\alpha$ between states with different values of
$J$. Only for a needle of vanishing diameter, $I_z/I_x \rightarrow
0$, which corresponds to $a \rightarrow 0$ in the case of an
ellipsoid, the transition is second order, see Fig.
\ref{Fig-EJJ_lambda=0}. It occurs at $\alpha = \left[1 -
1/(2S)^2\right]^{-1}$. This case is equivalent to the rotation
about a fixed axis studied in Ref.\ \onlinecite{chugar10prb}. For
any finite ratio $I_z/I_x$ the transition is first order, see Fig.
\ref{Fig-EJJ_lambda=2}. It occurs at the value of $\alpha$ that
depends on $I_z/I_x$. The origin of the transfer from a
second-order transition at $\lambda = 0$ to the first-order
transition at $\lambda \neq 0$ can be traced to the term $[J (J+1)
- K^2]S^{-2}\lambda$ in Eq.\ (\ref{eq:energy}). We should notice
that for a finite-size nanomagnet the analogy with first- and
second-order phase transition is, of course, just an analogy. To
talk about real phase transitions one has to take the limit of $S
\rightarrow \infty, I_{x,z} \rightarrow \infty$ when the distances
between quantum levels go to zero and the energy becomes
quasi-continuous function of $J$.

For a given $\lambda$, as $\alpha$ increases the ground state
switches from $J = 0$ to higher $J$ when
\begin{equation}
E_{00}^{(-)}( \alpha^0_J (\lambda) ) = E_{JJ}^{(-)} ( \alpha^0_J
(\lambda) )\,.
\end{equation}
Solution of this equation for $\alpha^0_ J ( \lambda )$ gives
%
\begin{equation}
\alpha^0_J = \frac{ (2 S)^2 (J + \lambda) }{ J [ (2 S)^2 - (J +
\lambda)^2 ] }\,. \label{eq:alpha_0J}
\end{equation}
This first transition occurs for the smallest value of $\alpha^0_J
(\lambda)$ and the transition is from $J = 0$ to the corresponding
critical value, $J_c$.  For $\alpha < \alpha^0_{J_c}$ the ground
state corresponds to $J = 0$ and $C_{\pm S} = 1$.  After the first
transition from $J = 0$ to $J = J_c$, the ground state switches to
sequentially higher $J$ at values of $\alpha$ which satisfy
\begin{equation}
E_{J-1 \; J-1}^{(-)}( \alpha_{J} (\lambda) ) = E_{JJ}^{(-)} (
\alpha_{J} (\lambda) )\,.
\end{equation}
Solution of this equation for $\alpha_{J} ( \lambda )$ gives
%
\begin{equation}
\alpha_J = \frac{ (2 S)^2 T(J, \lambda) }
    {  \sqrt{ (2 S)^2 (2 J - 1)^2 - T(J, \lambda)^2 } \sqrt{ (2 S)^2 - T(J, \lambda)^2 }
    }\,,
    \end{equation}
with
\begin{equation}
 T(J, \lambda) = 2 J - 1 + \lambda \label{eq:alpha_J}\,.
\end{equation}
The critical $\alpha_J$ has poles at $\lambda = 2 (S - J) + 1$.
For $\lambda \geq 1$ there is no longer a ground state transition
to $J = S$, even for very large values of $\alpha$.

\begin{figure}
\begin{center}
\includegraphics[scale=1]{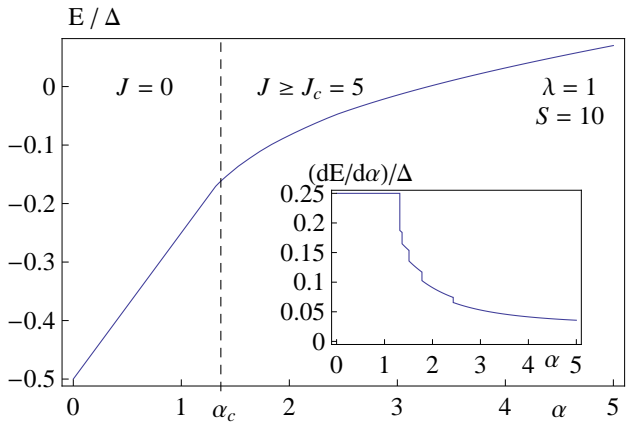}
\end{center}
\caption{ Dependence of the ground-state energy on $\alpha$ for a
spherical particle.  Inset shows the discontinuity of the
derivative of the ground state energy on $\alpha$.  }
\label{fig:sphere_gse}
\end{figure}
{\it Needle of vanishing diameter:} The case of a particle that
can only rotate about its anisotropy axis \cite{chugar10prb} is
equivalent in our model to a needle of vanishing diameter ($a
\rightarrow 0$ for an ellipsoid), having $\lambda = 0$. It is also
equivalent to the problem of tunneling of the angular momentum of
a superconducting current in a flux qubit coupled to a torsional
resonator. In this limit we reproduce results of Ref.
\onlinecite{chugar10prb}. The quantum number $K$ determines the
ground state, as the energy, Eq.~\eqref{eq:energy}, no longer
formally depends on $J$. However, the values of $\alpha_J$ at
$\lambda = 0$, for which ground state transitions occur, are the
same as those for which $E_{J \; K-1}^{(-)} = E_{J K}^{(-)}$, and
we will use $J$ to describe the ground state of the axial rotor as
well.  The first ground state transition occurs from $J = 0$ to $J
= 1$ at $\alpha(\lambda) = \alpha_1 (0) = \alpha^0_1 (0)$, because
$J_c = 1$ for $\lambda \lesssim 0.01$. At $\alpha = \alpha_2 (0)$
the ground state switches from $J = 1$ to $J = 2$, and so on.  The
final transition is to a completely localized spin state $J = S$
in which spin tunneling is frozen for all $\alpha > \alpha_S (0)$.
For example, when $S = 10$, $\alpha_1 (0) = \alpha^0_1 (0) =
1.0025$ and $\alpha_{10} = 3.2066$.

\begin{figure}
    \includegraphics[trim = 2cm 0cm 2cm 0cm, scale=0.68]{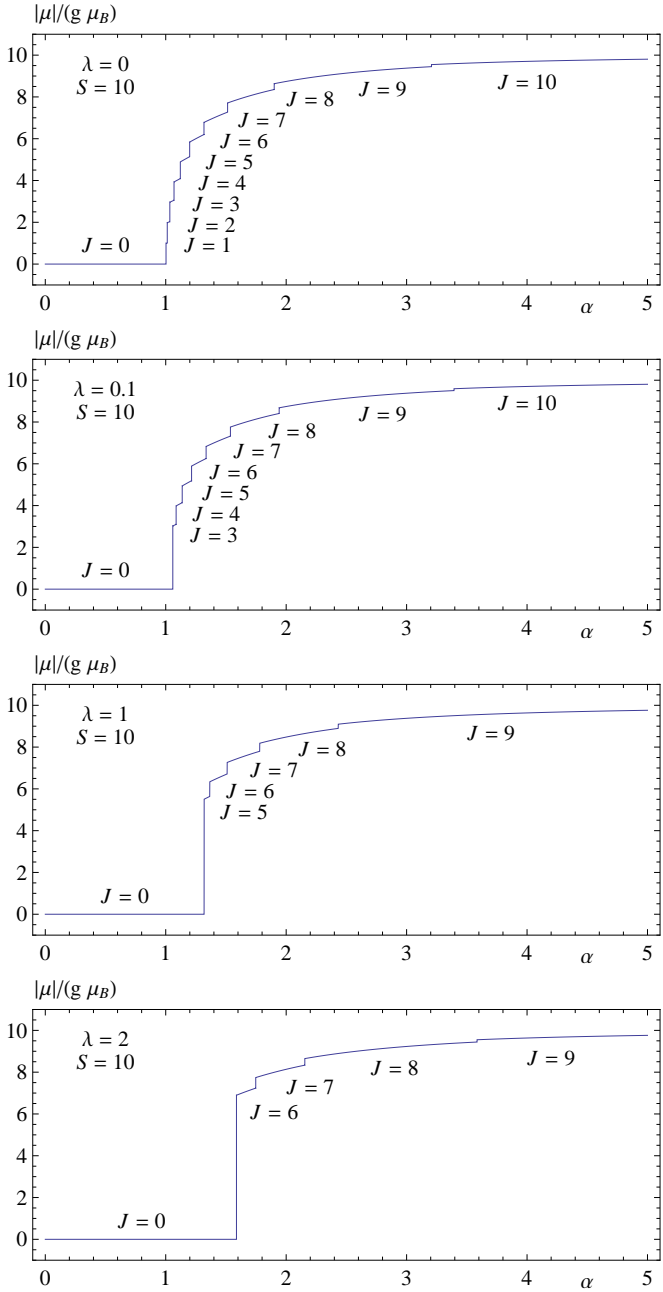}
\caption{ Ground state magnetic moment for a needle of vanishing
diameter $(\lambda = 0)$, finite-diameter needle $(\lambda =
0.1)$, sphere $(\lambda = 1)$, and a disk of vanishing thickness
$(\lambda = 2)$.} \label{fig:moments}
\end{figure}
{\it Needle of finite diameter:} The ground state of a needle of
finite diameter ($a \ll c$ for an ellipsoid) with $\lambda = 0.1$,
that is free to rotate about any axis, shows qualitatively
different behavior. As $\alpha$ increases, the ground state
changes from $J = 0$ to $J_c = 3$ at $\alpha = \alpha^0_3 (0.1)$,
as the smallest value of $\alpha^0_J (0.1)$ for $1 \leq J \leq S$
occurs for $J_c = 3$. The $J = 1, 2$ states never become the
ground state.  After this, transitions occur to successively
higher $J$, beginning with $J = 4$ at $\alpha = \alpha_4 (0.1)$,
and eventually localizing the spin with $J = S$ for $\alpha >
\alpha_S (0.1)$.  For $S = 10$, $\alpha^0_3 (0.1) = 1.0588$ and
$\alpha_{10} (0.1) = 3.3935$.

{\it Sphere:} As $\lambda$ increases towards unity, the particle
becomes more symmetric with the moment of inertia having (prolate)
ellipsoidal symmetry, until it reaches spherical symmetry at
$\lambda = 1$. The first ground state transition occurs from $J =
0$ to $J = J_c = 5$ at $\alpha = \alpha^0_5 (1)$, and subsequent
transitions occur at $\alpha = \alpha_J (1)$.  However, the spin
never localizes in the $J = S$ state even for very large alpha, as
$\alpha_J (1)$ has a pole at $J = S$, so the last transition
occurs to the $J = S - 1$ state at $\alpha = \alpha_{S - 1} (1)$.
For $S = 10$, $\alpha^0_5 (1) = 1.3187$ and $\alpha_9 (1) =
2.4325$.

{\it Disk:} With $\lambda$ increasing from unity, the symmetry of
the body becomes that of an oblate ellipsoid, and begins to
flatten in the plane perpendicular to the anisotropy axis. It is
easy to check from Eq.\ (\ref{eq:energy}) that for $1 < \lambda
\leq 2$ the state with $J = S$ always has higher energy than the
state with $J = S-1$, even in the limit of $\alpha \rightarrow
\infty$. This means that for an oblate particle some spin
tunneling (accompanied by mechanical rotations) survives in the
ground state no matter how light the particle is. This purely
quantum-mechanical result has no semi-classical analogy. In the
case of a disk of vanishing thickness, $\lambda = 2$, the first
ground state transition occurs from $J = 0$ to $J = J_c = 6$ at
$\alpha = \alpha^0_6 (2)$, and subsequent transitions occur at
$\alpha = \alpha_J (2)$ up through $J = S - 1$.  For $S = 10$,
$\alpha^0_6 (2) = 1.5873$ and $\alpha_9 (2) = 3.5849$.

\section{Ground-state magnetic moment \label{moment}}
%
As has been already mentioned, the magnetic moment is due entirely
to the spin of the particle, as $L_z$ represents mechanical motion
of the particle as a whole, and not electronic orbital angular
momentum. Thus,
%
\begin{equation}
\mu = - g \mu_B \bra{\Psi_{JK}} S_z \ket{\Psi_{JK}} = - g \mu_B S
    \frac{\alpha K}{\sqrt{S^2 + (\alpha K)^2}}\,.
\end{equation}
Here $g$ is the spin gyromagnetic factor, and the minus sign
reflects the negative gyromagnetic ratio $\gamma = - g \mu_B /
\hbar$.  The ground state always corresponds to $J = K$, so these
are used interchangeably in descriptions of the ground state.

The dependence of the magnetic moment on $\alpha$ for different
aspect ratios of the particle is shown in Fig. \ref{fig:moments}.
For $\alpha < \alpha_{J_c} (\lambda)$ the ground state corresponds
to $J = K= 0$, so the spin-up and spin-down states are in an equal
superposition which produces zero magnetic moment.  At greater
values of $\alpha$ the spin states contribute in unequal amounts
which leads to a non-zero magnetic moment.  As $\alpha$ becomes
large, the magnetic moment approaches its maximal value $
|\mu_{max}| = g \mu_B S$. Note that the magnetic moment approaches
its maximum value even for values of $\lambda$ that do not admit
transitions to $J = S$ states.
\begin{figure}
\begin{center}
\includegraphics[trim = 1cm 0cm 0cm 0cm, scale=0.56]{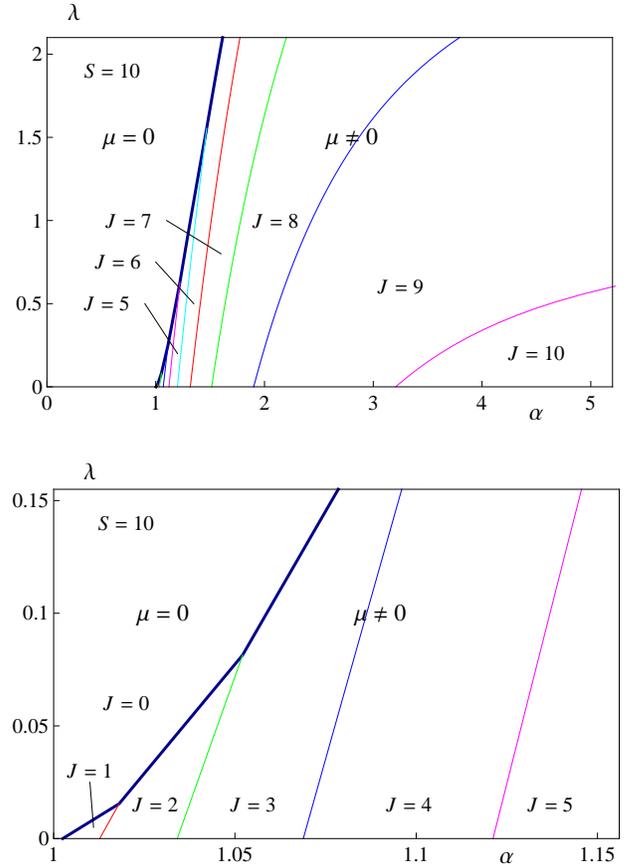}
\end{center}
\caption{ Quantum phase diagram for the ground-state magnetic
moment and the total angular momentum. } \label{fig:phase_diagram}
\end{figure}

%
%
%
%
%
%
%

Because the ground state is completely determined by the
parameters $\alpha$ and $\lambda$, we can depict the ground state
behavior in a quantum phase diagram shown in Fig.
\ref{fig:phase_diagram}. The curves separate areas in the
(${\alpha},\lambda$) plane that correspond to different values of
$J$ and different values of the magnetic moment. Notice the fine
structure of the diagram (lower picture in Fig.
\ref{fig:phase_diagram}) near the first critical $\alpha$. This
very rich behavior of the ground state on parameters must have
significant implications for magnetism of rigid atomic clusters.
%
%
%
%
%
%


\section{Conclusions \label{conclusions}}

We have studied the problem of a quantum rotator containing a
tunneling spin. This problem is relevant to quantum mechanics of
free magnetic nanoparticles. It also provides an interesting
insight into quantum mechanics of molecules studied from the
macroscopic end. The answer obtained for the energy levels of a
symmetric rotator, Eq.\ (\ref{eigenstates-EJK}), is
non-perturbative and highly non-trivial. It is difficult to
imagine how it could be obtained from first principles without the
reduction to two spin states. Indeed, for spin $S$ the tunnel
splitting itself generally appears in the $S$-th order of
perturbation theory, see Eq.\ (\ref{Delta}), so the path from the
full crystal-field Hamiltonian like, e.g., Eq.\ (\ref{CF}) to Eq.\
(\ref{eigenstates-EJK}) must be very long. Equations
(\ref{eq:eigenstates}) and (\ref{eigenstates-EJK}) represent,
therefore, a unique exact solution of the quantum-mechanical
problem of a mechanical rotator with a spin. Striking feature of
this solution is presence of  first- and second-order quantum
phase transitions between states with different values of the
magnetic moment.

Our results provide the framework for comparison between theory
and experiment on very small free magnetic clusters. Our main
conclusion for experiment is that rotational states and magnetic
moments of such clusters depend crucially and in a predictable way
on size and aspect ratio. This dependence results in a complex
phase diagram that separates regions in the parameter space,
corresponding to different values of the magnetic moment. Broad
distribution of the magnetic moments that does not simply scale
with the volume, has, in fact, been reported in beams of free
atomic clusters of ferromagnetic materials \cite{Billas,Xu,Payne}.
Our results may shed some additional light on these experiments.
They may also apply to free magnetic molecules if one can justify
the condition of rigidity. Direct comparison between theory and
experiment may be possible for atomic clusters (molecules) in
magnetic traps.

To see that the quantum problem studied in this paper may, indeed,
be relevant to quantum states of free nanomagnets, consider, e.g.,
a spherical atomic cluster of radius $R$ and average mass density
$\rho$ having spin $S = 10$ that, when embedded in a large body,
can tunnel between up and down at a frequency of a few GHz, thus
providing $\Delta \sim 0.1$K. Significant changes in the magnetic
moment of such a cluster would occur at $\alpha \sim 1$, which,
according to Eq.\ (\ref{eq:alpha}), corresponds to $I = 8\pi \rho
R^5/15 \sim 10^{-42}$kg\,m$^2$ and $R \sim 1$nm. For a magnetic
molecule like, e.g., Mn$_{12}$, the moments of inertia would also
be in the ballpark of $10^{-42}$kg\,m$^2$. However, the natural
spin tunnel splitting in Mn$_{12}$ is very small, thus, providing
a very large $\alpha$. Same is true for Fe$_8$ magnetic molecules.
In this case the spin tunneling in a free molecule must be
completely frozen. Even if the molecule cannot be considered as
entirely rigid, such effect, if observed, would receive natural
interpretation within the framework of our theory.\\

\section{Acknowledgements}
This work has been supported by the U.S. Department of Energy
through grant No. DE-FG02-93ER45487.

\end{document}